\documentclass[preprint,showpacs,preprintnumbers,amsmath,amssymb]{revtex4}
\usepackage{graphicx}
\begin{document}
\title{Plasmon Excitations for Encapsulated Graphene       }
\author{ Godfrey Gumbs$^{1,2}$,  N. J. M. Horing$^3$, Andrii Iurov$^{ 4}$, and Dipendra Dahal$^{1}$}
\affiliation{$^{1}$Department of Physics and Astronomy, Hunter College of the
City University of New York, 695 Park Avenue, New York, NY 10065, USA\\
$^{2}$ Donostia International Physics Center (DIPC),
P de Manuel Lardizabal, 4, 20018 San Sebastian, Basque Country, Spain\\
$^{3}$ Department of Physics and Engineering Physics, Stevens Institute of Technology, Hoboken, NJ 07030, USA \\
$^{4}$Center for High Technology Materials, University of New Mexico, NM 87106, USA}

\date{\today}

\begin{abstract}

We have developed  an analytical formulation to calculate the plasmon dispersion
relation for a two-dimensional layer which is encapsulated within  a narrow spatial
   gap between two bulk half-space plasmas. This is based on a solution  of
   the inverse dielectric function integral equation within the random-phase
   approximation (RPA).  We take into account  the
   nonlocality  of the plasmon dispersion relation for both   gapped and
   gapless graphene as the sandwiched two-dimensional (2D) semiconductor
   plasma. The associated nonlocal graphene plasmon
spectrum coupled to the ``sandwich" system is exhibited in density plots, which show a linear
mode and a pair of depolarization modes shifted from the bulk plasma frequency.

\end{abstract}

\vskip 0.2in

\pacs{73.21.-b, 71.70.Ej, 73.20.Mf, 71.45.Gm, 71.10.Ca, 81.05.ue}

\maketitle\section{Introduction}
\label{sec1}

The properties of high-quality graphene encapsulated between two films such as
hexagonal boron-nitride are just beginning to be explored and have now become an
active area of research due to recent advances in  device fabrication techniques
\cite{encaps1,encaps2,encaps3,encaps4,encaps5,encaps6,encaps7,encaps8,encaps9,encaps10}.
Interest in the optical properties of these heterostructures has been focused on their
unusual plasmonic behavior including their  spatial  dispersion and damping.
This hybrid system may be employed for tailoring novel  metamaterials.
Additionally, this fabrication technique provides a clean environment for graphene.
This brand new area  of nanoscience not only poses  challenges for experimentalists, but also
for theoreticians seeking to formulate a theory for a model system. Although
 there already exists a copious literature on graphene on a single substrate
 \cite{GG,ONB,NJMH,Pol1,Pol2,Pol3,Pol4,Pol5},  this  study shows that
  encapsulated graphene is vastly different in many ways.

\medskip
\par

\begin{figure}
\centering
\includegraphics[width=0.35\textwidth]{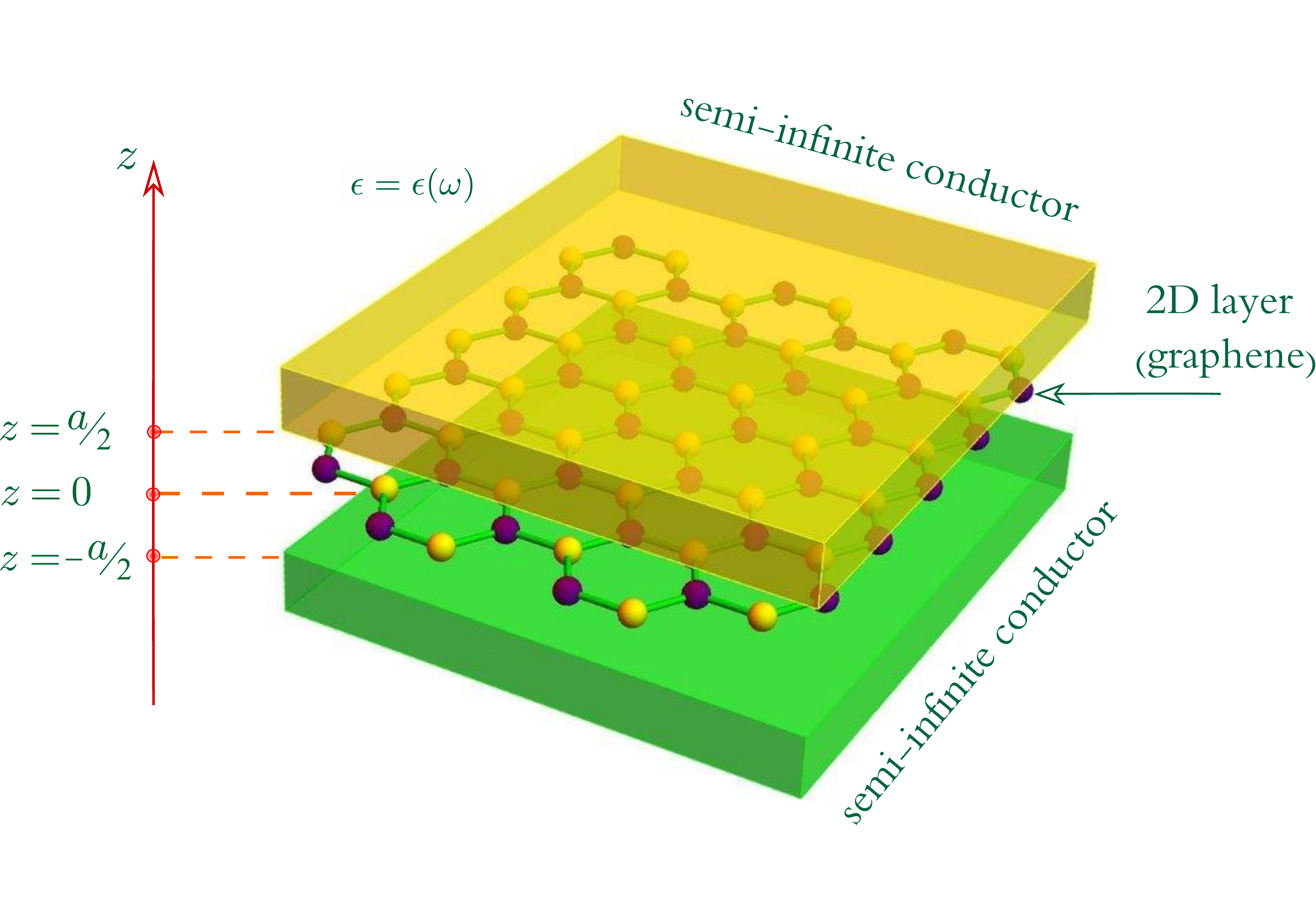}
\caption{(Color online) Schematic illustration of a pair of thick conducting plasmas
(taken to be semi-infinite in our formulation of the problem) encapsulating a 2D
monolayer graphene sheet in a ``sandwich" array.}
\label{FIG:1}
\end{figure}

The model we use in this paper consists of
 two identical semi-infinite metallic plasmas with planar  boundaries at $z = \pm a/2$.
  Within the  spatial separation between the two  bulk conducting plasmas ($|z|<a/2$) is inserted
   a 2D monolayer graphene sheet  at $z=0$, shown schematically in Fig. \ref{FIG:1}. The natural first step in our calculations of the plasmon excitation spectrum is to  set up and solve  the  random-phase approximation  (RPA) integral equation for the inverse dielectric screening function  of this hybrid system.
 We have   solved this equation  analytically in position representation for a narrow spatial
   gap between the bulk half-space plasmas, obtaining a closed-form formula for the inverse
   dielectric  function in terms of the nonlocal  polarizability for graphene and the bulk
   metallic polarizability, for which the latter is well approximated by the   hydrodynamical
   model.  Based on this newly derived formula, we have calculated the nonlocal plasmon
   dispersion relation numerically, considering both
   gapped and gapless graphene as the two-dimensional
	(2D) semiconductor plasma.  The resulting nonlocal graphene plasmon spectra coupled to the
	``sandwich" system are exhibited in density plots, which show a linear mode and a pair of  depolarization  modes shifted from the bulk plasma frequency.

\medskip
\par

 Hexagonal boron nitride has been the main substrate material that facilitates
 graphene based devices to exhibit micrometer-scale
ballistic transport. The recent work of Kretinin, et al. \cite{encaps10}
has shown that other atomically flat crystals may also be
employed as substrates for making high-quality graphene heterostructures.
Alternative substrates  for  encapsulating graphene include molybdenum and
tungsten disulfides which have been   found to exhibit consistently
high carrier mobilities of about $6\times 10^4$ cm$^2$  V$^{-1}$ s$^{-1}$. On the other hand,
when graphene is encapsulated with atomically flat layered oxides such as mica,
bismuth strontium calcium copper oxide, or vanadium pentoxide,
the result is remarkably low quality  graphene  with
mobilities of about $  10^3$ cm$^2$  V$^{-1}$ s$^{-1}$. This difference
is due  mainly to self-cleansing which occurs  at interfaces
between graphene, hexagonal boron nitride, and transition metal dichalcogenides.
 In our model calculations, we allow for the possibility that the substrate may
  affect the energy  band structure of graphene by opening a gap in the energy band.
  We compare the resulting calculated plasmon spectra for encapsulated gapless and  gapped graphene.

The work of Chuang, et al. \cite{DD1}  is a study of the graphene-like
 high mobility  for  p- and n-doped  WSe$_2$. Electron transport
in the junction between two 2D  materials $MoS_2$ and graphene have also
been reported recently, showing how spatial confinement  can influence
physical properties \cite{DD2}.
Mobility, transconductance   and carrier inhomogeniety experiments  have
also been reported  in \cite{DD3,DD4,DD5,DD6}
for monolayer and bilayer graphene fabricated on    hexagonal BN   as well as mica based
substrates.

\medskip
\par

The outline of the rest of our paper is as follows.  In Sec. \ \ref{sec2},
we give details of our calculation of the inverse dielectric function for
a 2D layer sandwiched between two conducting substrates whose separation is
very small. We explicitly derive the plasma dispersion equation when the thick
substrate layers may be treated in the hydrodynamical model.
The 2D RPA ring diagram polarization function for graphene at arbitrary temperature
is employed in the dispersion  equation. A careful determination of the plasmon spectra
for a range of energy gap and carrier doping values is reported in Sec.\ \ref{sec3}.
We conclude with a discussion of the highlights of our calculations in Sec.\ \ref{sec4}.

\section{  Theoretical   Formulation}
\label{sec2}

We consider two identical semi-infinite conductors on either side of a 2D
semiconductor layer (Fig.\ \ref{FIG:1}).  One
of the conductors  extends from $z=-a/2$ to $z=-\infty$ while the other
conductor has its surface at $z= a/2$ and extends to $z=\infty$.  The 2D
layer  lies in a plane mid-way between the two conductors at $z=0$
in the gap region $|z|<a/2$.
The  inverse dielectric function satisfies

\begin{eqnarray}
K(z_1,z_2)&=& K_\infty(z_1,z_2)
\nonumber\\
&-& \int_{-\infty}^\infty dz^{\prime}  \int_{-\infty}^\infty dz^{\prime\prime} \
K_\infty (z-z^\prime) \left[\alpha_{2D}(z^\prime,z^{\prime\prime})- \alpha_{gap}(z^\prime,z^{\prime\prime})
\right] K(z^{\prime\prime},z_2)\ ,
\label{eq:1}
\end{eqnarray}where $K_\infty(z_1,z_2)=K_\infty(z_1-z_2)$ is the bulk infinite space symmetric conducting
medium inverse dielectric function and

\begin{equation}
\alpha_{2D}(z^\prime,z^{\prime\prime})=\frac{2\pi e^2}{\epsilon_s q_\parallel}
\Pi^{(0)}_{2D} (q_\parallel,\omega) e^{-q_\parallel |z^\prime|} \delta(z^{\prime\prime})
\equiv \tilde{\alpha}_{2D}(q_\parallel,\omega) \ e^{-q_\parallel |z^\prime|} \delta(z^{\prime\prime}) \ ,
\label{eq:2}
\end{equation}
which defines $\tilde{\alpha}_{2D}(q_\parallel,\omega) $ in terms of the 2D ring diagram
$\Pi^{(0)}_{2D} (q_\parallel,\omega)$. For a narrow gap between the two identical half-space
slabs, i.e., $q_\parallel a \ll 1$, the gap polarizability is given by

\begin{equation}
\alpha_{gap}(z^\prime-z^{\prime\prime})= a\delta(z^{\prime\prime}) \alpha_{\infty}(z^\prime-z^{\prime\prime})
= a\delta(z^{\prime\prime}) \int_{-\infty}^\infty \frac{dp_z}{2\pi}\alpha_\infty (p_z,q_\parallel)
   e^{ip_z  (  z^\prime-z^{\prime\prime}  )  } \ .
\label{eq:3}
\end{equation}

Employing Eqs. \ (\ref{eq:2}) and (\ref{eq:3}) in Eq. \ (\ref{eq:1}), we have
$K(z_1,z_2)= K_\infty(z_1-z_2)- {\cal F}(z_1;q_\parallel,\omega)    K(0,z_2)$,
where ${\cal F}(z_1;q_\parallel,\omega)= {\cal F}_{2D}(z_1;q_\parallel,\omega)
+ {\cal F}_{gap}(z_1;q_\parallel,\omega) $
and

\begin{eqnarray}
{\cal F}_{2D}(z_1;q_\parallel,\omega)&\equiv & \frac{2\pi e^2}{\epsilon_s q_\parallel}
\Pi^{(0)}_{2D} (q_\parallel,\omega)
\int_{-\infty}^\infty dz^\prime\ K_\infty (z_1-z^\prime)
   e^{-q_\parallel |z^\prime|} \ ,
	\nonumber\\
	{\cal F}_{gap}(z_1;q_\parallel,\omega)&\equiv &
-a\  \int_{-\infty}^\infty dz^\prime\ K_\infty (z_1-z^\prime)  \alpha_\infty(z^\prime,0) \ .
\label{eq:5-2}
\end{eqnarray}

Setting $z_1=0$  in the equation relating $K(0,z_2)$ to $K(z_1,z_2)$ above,
we may solve for $K(0,z_2)$ and then obtain

\begin{equation}
K(z_1,z_2)= K_\infty(z_1-z_2)- \frac{{\cal F}(z_1;q_\parallel,\omega)}
{1+{\cal F}(z_1=0;q_\parallel,\omega)}    K_\infty(0,z_2)\ ,
\label{eq:6}
\end{equation}
so that the plasma  excitation frequencies are determined by solving for the zeros of

\begin{eqnarray}
{\cal D}(q_\parallel,\omega) &\equiv & 1+{\cal F}(z_1=0;q_\parallel,\omega)
\nonumber\\
&=& 1+ \int_{-\infty}^\infty dz^\prime\ K_\infty (0-z^\prime))
\left[  \frac{2\pi e^2}{\epsilon_s q_\parallel}  \Pi^{(0)}_{2D} (q_\parallel,\omega) e^{-q_\parallel |z^\prime|}
-a\  \alpha_\infty(z^\prime,0)\right] \ .
\label{eq:7}
\end{eqnarray}

Expressing Eq.   (\ref{eq:5-2}) in terms of the Fourier transform of
$K_\infty (z_1-z^\prime)$, we obtain

\begin{eqnarray}
{\cal F}_{2D}(z_1;q_\parallel,\omega)&= &  \frac{2  e^2}{\epsilon_s}
\Pi^{(0)}_{2D} (q_\parallel,\omega)
\int_{-\infty}^\infty dp_z\ e^{ip_zz_1}  \frac{K_\infty (p_z )}{p_z^2+q_\parallel^2} \ ,
	\nonumber\\
	{\cal F}_{gap}(z_1;q_\parallel,\omega)&= &
-a\  \int_{-\infty}^\infty \frac{dp_z}{2\pi}\  e^{ip_zz_1}  K_\infty (p_z) \alpha_\infty(p_z) \ .
\label{eq:5-3}
\end{eqnarray}

Using the hydrodynamical model for nonlocality of the conducting substrate \cite{EGUI,Kamen}, we have
$\epsilon(q,\omega)=1-\omega_p^2/(\omega^2-\beta^2q^2)$, where $\omega_p$ is the
local bulk metallic plasma frequency, $\beta$ is an adjustable parameter which mabe adjusted to give the correct dispersive shift of the bulk  plasma frequency, i.e.,
$\beta^2=3v_F^{(3D)\ 2}/5$ (where $v_F^{(3D) }$ is the bulk Fermi velocity of the conducting substrate).
This approximation   leads to the dispersion equation which,
after performing the $p_z$ integrations, we obtain

\begin{eqnarray}
{\cal D}(q_\parallel,\omega)&=&1+  \frac{2 \pi  e^2}{\epsilon_s q_\parallel}
\Pi^{(0)}_{2D} (q_\parallel,\omega) \frac{1}{|A(q_\parallel,\omega)|}
\left( \frac{\omega_p^2 q_\parallel -\omega^2 |A(q_\parallel,\omega)|}{\omega_p^2-\omega^2}  \right)
\nonumber\\
&-&   \frac{  a \omega_p^2}{2\beta^2 |A(q_\parallel,\omega)|}\ ,
\label{xxx}
\end{eqnarray}
where
$A(q_\parallel,\omega) \equiv  \beta ^{-1} \left(  \omega_p^2+\beta^2 q_\parallel^2
-\omega^2\right)^{1/2} $.

The 2D RPA ring diagram polarization function for graphene with a gap $\Delta$  may be expressed as

\begin{eqnarray}
\label{A1}
\Pi^{(0)}_{2D}(q,\omega)
&& = \frac{g}{4 \pi^2} \int d^2 {\bf k} \sum\limits_{s,s' = \pm} \left( 1 + s s'
\frac{{\bf k} \cdot ({\bf k}+{\bf q}_{\parallel}) + \Delta^2}{\epsilon_k \,\, \epsilon_{\vert {\bf k}+{ \bf q}_{\parallel} \vert }}  \right) \nonumber \\ && \times \frac{f(s \, \epsilon_{{\bf k}}) - f(s' \epsilon_{{\bf k}+{\bf q}_{\parallel}})}{s \, \epsilon_{{\bf k}_{\parallel}} - s' \epsilon_{{\bf k}+{\bf q}_{\parallel}} - \hbar \omega - i \eta^+ } \ ,
\end{eqnarray}
where $s, s^\prime$ are band indices, $g=4$ accounts for both spin and valley degeneracies and
$\epsilon_{{\bf k}}$  is the band energy. Since we limit our considerations to zero temperature,
$T=0$, the Fermi-Dirac distribution function is reduced to the Heaviside step function
$f(\epsilon, \mu; T \rightarrow 0) = \eta_{+}(\mu - \epsilon)$, and
we use the results of Refs.  [\onlinecite{Wunsch,pavlo}].

%%%%  Pol5,

\section{Numerical Results and Discussion}
\label{sec3}

\begin{figure}
\centering
\includegraphics[width=0.55\textwidth]{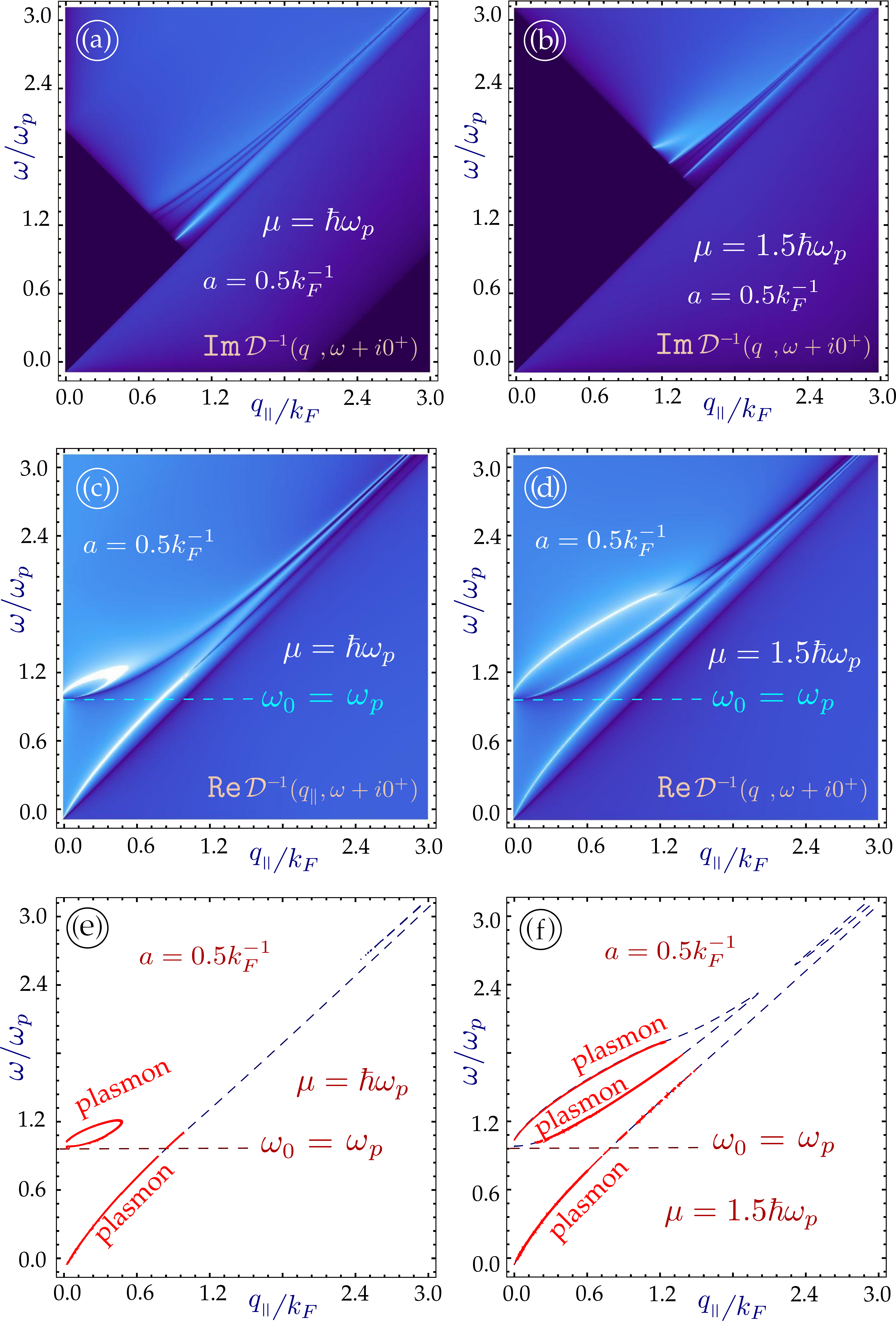}
\caption{(Color online) Particle-hole modes and plasmon dispersion for doped, gapless
graphene sandwiched between two semi-infinite conducting plasmas with separation
$a=0.5\ k_F^{-1}$. Panels $(a)$ and $(b)$ present  the regions  where the plasmons
are damped for $\mu=\hbar\omega_p$  and $\mu=1.5\hbar\omega_p$.\, respectively. Panels
$(c)$ and $(d)$ show density plots for the plasmon excitation spectra corresponding to
the  parameters for zero  energy gap  and the same bulk plasma separation chosen in $(a)$ and
$(b)$. Panels $(e)$ and $(f)$ show the plasmon excitations, both damped (dashed lines)
and undamped   (solid lines), obtained by solving $ \mbox{Re} \, D (q_\parallel,\omega) = 0 $
for the chosen distance   $a=0.5k_F^{-1}$ between
the bulk surfaces with the 2D layer at $z=0$, the same as in  $(a)$ through $(d)$. The value of the parameter
in the hydrodynamical model is $\beta=c_0 v_F$   in terms of the Fermi velocity $v_F$ for graphene
and $c_0=(3/5)^{1/2}(v_F^{3D}/v_F)$, where $v_F^{3D}$ is the Fermi velocity of the conducting substrate
(but we take  $c_0=1$ here).}
\label{FIG:2}
\end{figure}

\begin{figure}
\centering
\includegraphics[width=0.55\textwidth]{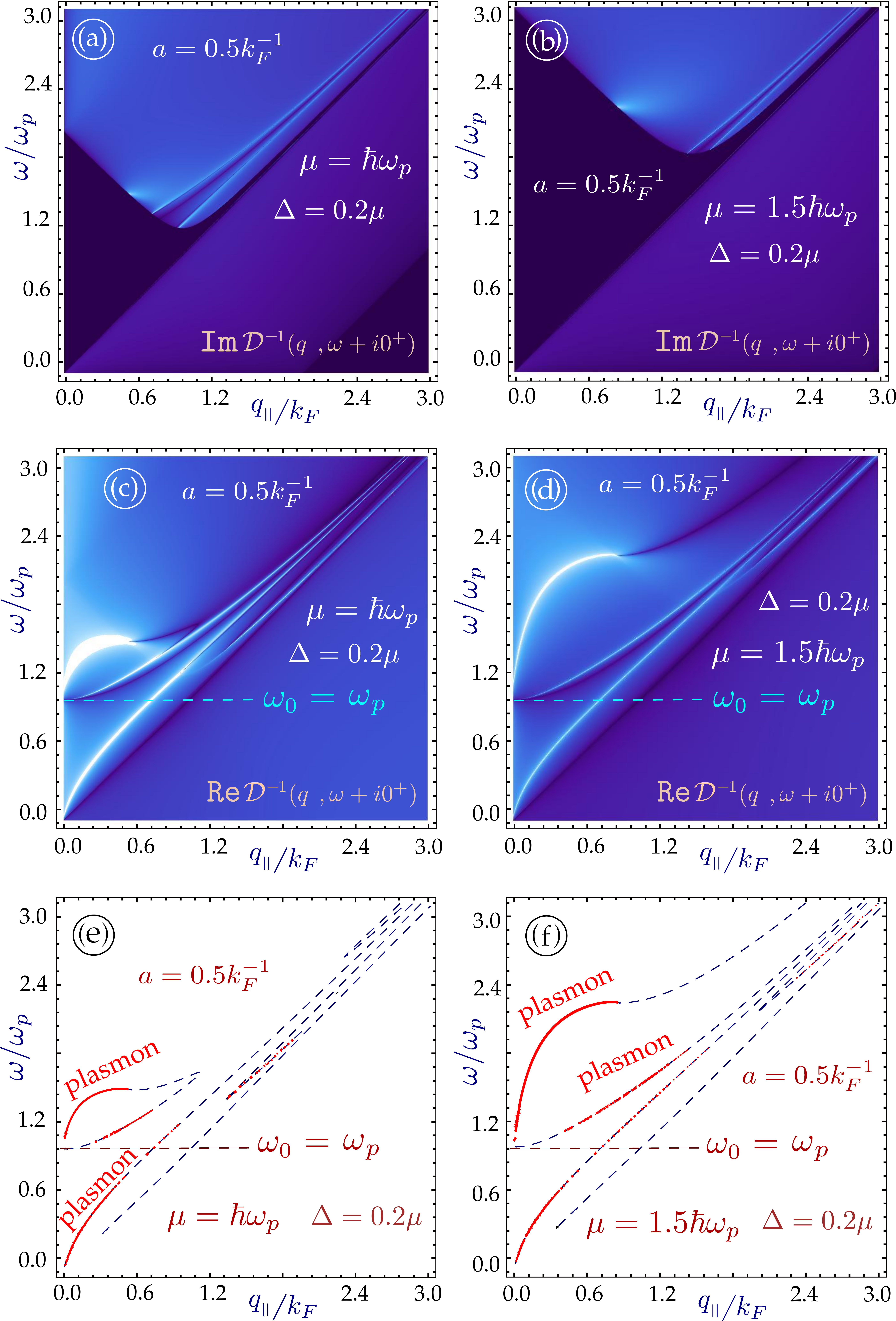}
\caption{(Color online) Particle-hole modes and plasmon dispersion for doped graphene
encapsulated between two semi-infinite conducting plasmas with separation $a=0.5\ k_F^{-1}$.
The graphene layer possesses an intermediate value for the  energy bandgap $\Delta=0.2\ \mu$
 determining  the energy dispersion.  Panels $(a)$ and $(b)$ present density plots for
 $ \mbox{Im}\, D^{-1} (q_\parallel,\omega)$, showing the regions  where the plasmons are
 damped when $\mu=\hbar\omega_p$  and $\mu=1.5\hbar\omega_p$,   respectively. Panels $(c)$
 and $(d)$ exhibit density plots for the plasmon excitation spectra corresponding
to  the  parameters for the  energy gap $\Delta=0.2\mu$ and bulk plasma separation chosen in $(a)$ and
$(b)$. Panels $(e)$ and $(f)$ show the plasmon excitations, both damped (dashed lines) and undamped (solid lines),
obtained by solving $ \mbox{Re} \, D (q_\parallel,\omega) = 0 $ for the  chosen  distance between
the bulk surfaces with the 2D layer at $z=0$, the same as $(a)$ through $(d)$. The value of the parameter
used in the hydrodynamical model is $\beta=v_F$ in terms of the Fermi velocity for graphene,
as in Fig.\ \ref{FIG:2}.}
\label{FIG:3}
\end{figure}

\begin{figure}
\centering
\includegraphics[width=0.55\textwidth]{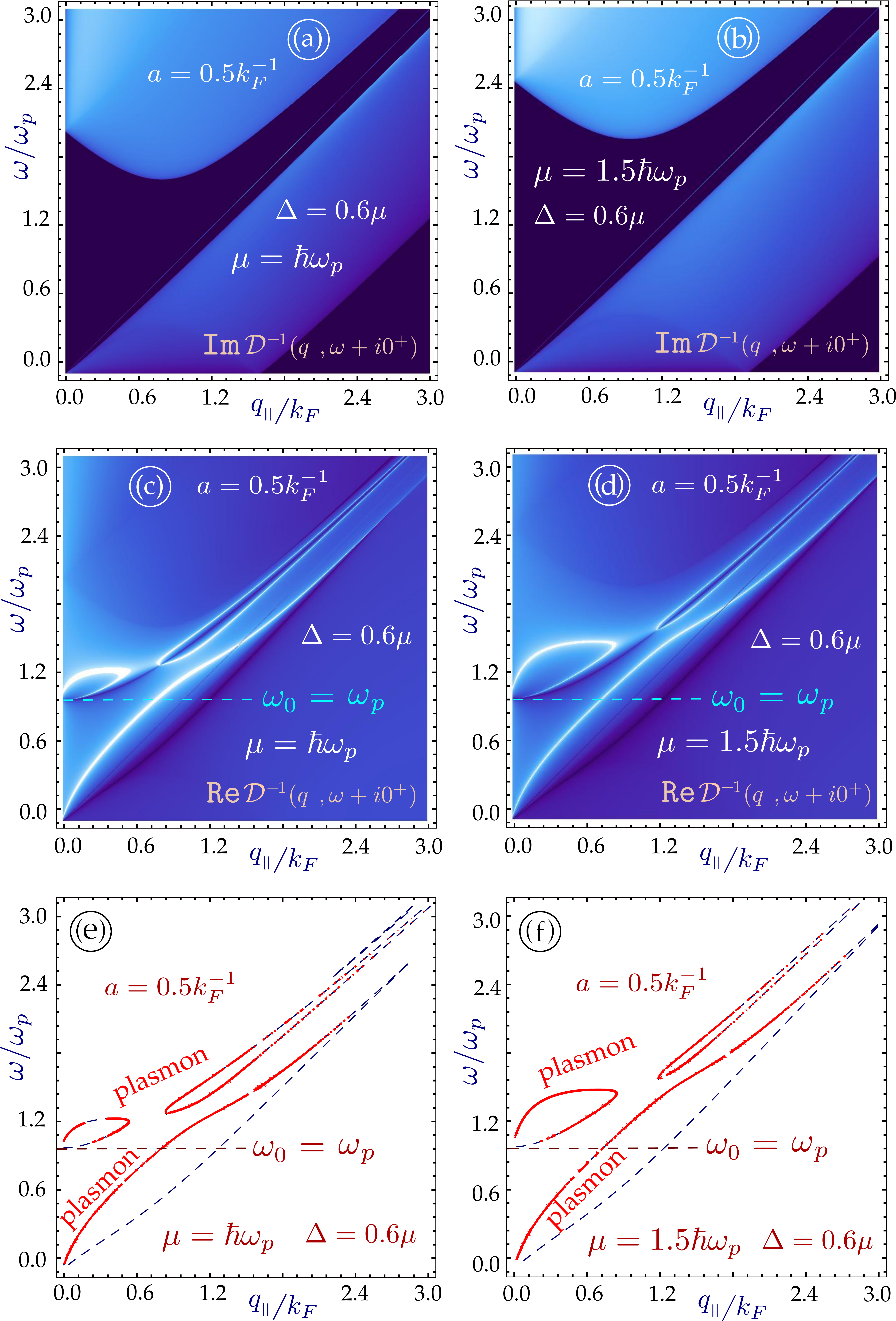}
\caption{(Color online) Particle-hole modes and the plasmons for graphene encapsulated
between two semi-infinite conducting substrates. Here, we  present a set of graphs,
 similar to those in Fig. \ref{FIG:3}, but for the case of a large energy gap $\Delta=0.6\ \mu$.}
\label{FIG:4}
\end{figure}

In the figures, we present calculated results  for the nonlocal  plasmon excitations of encapsulated
gapless and gapped graphene.  As we demonstrated  in previous work \cite{GG,ONB}, the hybrid
plasmon modes and their damping are mainly determined by the doping concentration, i.e., the
 chemical potential $\mu$, along with the energy bandgap $\Delta$. Consequently, we have paid particular
 attention  in our numerical investigations to the various regimes of the ratio
  $\Delta/\mu$.   In Fig.\ \ref{FIG:2}, we exhibit results for gapless graphene.
		Figures \ref{FIG:3} and \ref{FIG:4} illustrate the case in which the graphene
  layer has an intermediate or large energy gap. The regions of strong damping  arise from inter-band
transitions  which are forbidden by Pauli blocking (i.e., when the final transition states are filled,
so that further transitions cannot occur). In the case of a {\em single\/} substrate and a  2D layer,
 the lowest acoustic branch vanishes due to damping in the long wavelength limit for a range of
 separations between the 2D layer and the surface. However, this is not the present case under
consideration  involving encapsulation of the graphene sheet within a small gap between the two surfaces.
Within the ``small gap" framework, $a<1/q_\parallel$, the lower branch is never damped  in the long
 wavelength limit.    This result is quite   different  from that obtained for a  {\em single\/} substrate,
  in which strong damping of the acoustic branch appears in the long wavelength limit
   for the distance $a$   below a specific critical value. In regard to the two upper branches,
attributed to  bulk three-dimensional  plasmons,   their undamped parts
appear as upper and lower arcs of a single loop on the left sides of  the figures
in the cases of low doping or low   value of the chemical potential,
as clearly  demonstrated in Fig.\ref{FIG:2}. It is noteworthy that both bulk plasmon modes
start from $\omega_p$, {\em not\/} from $\omega_p/\sqrt{2}$ as we  previously observed in the case of a
single conducting substrate.

\par
\medskip
\par

The presence of a bandgap in the graphene energy spectrum leads to interesting features.
 First, a finite  energy gap modifies the location of each
single-particle excitation region. It is especially unusual to observe the extension of the upper
plasmon branches, which are understood to   arise  from a semi-infinite substrate. It is also apparent
that the acoustic mode is broken into two separate undamped parts located between the two distinct
single-particle excitation regions,   as has been previously      reported for free-standing gapped graphene
\cite{pavlo} (see Figs. \ref{FIG:3} (c) and (d)). The figures show that specifically acoustic
plasmon mode behavior persists in our present case of encapsulated graphene.
Within the framework of our assumption that the spacing between  the conducting half-spaces
is narrow, $a\ll 1/q_\parallel$,  the results are relatively insensitive to the gap separation,
and we consider  only small values of $a$. Moreover, we examined modification of the plasmon spectra
for various values of the Fermi energy, $\mu$,
as depicted  in Figs.\ \ref{FIG:2} through \ref{FIG:4}.

\medskip
\par

In the matter of   experimental realization of our results, it is necessary  to achieve a situation
in which  the surface plasmons are not Landau damped, i.e., the corresponding branches are located outside
of the upper inter-band part of the single-particle excitation spectrum. Accordingly, the frequency
of the surface plasmon mode should be comparable with the Fermi energy in graphene, as well as with the
 corresponding $q^{1/2}$-2D-graphene plasmon energy. Only if this condition is satisfied
  will it be possible to observe the strong plasmon coupling we have reported here.
Therefore, one must ensure that the bulk plasmon frequency is in the range of $\sim .1 \, eV$
and below \cite{NPo1}.   Ref [\onlinecite{NPo1}] is an experimental paper on  $Bi_2Se_3$ which is a heavily doped topological
insulator,   where the surface plasmon energy was found to be   around $0.1 \, eV$. Evidence of mutual
interaction between the surface plasmon and the Dirac plasmon of  $Bi_2Se_3$ has been provided by using
high-resolution electron energy loss spectroscopy.  Additionally, at a graphene/$Bi_2Se_3$ interface,
which was recently experimentally realized by Kepaptsoglou, et al. \cite{NPo2}, the surface plasmon of
$Bi_2Se_3$  is hybridized with acoustic plasmons in graphene.  In this vein, the Fermi   energy of
free-standing graphene  corresponding to an electron density $n=10^{16}\, m^{-2}$
is     $  E _F = \hbar v_F \sqrt{\pi n } = 0.12\, eV$, so  the two
quantities   are of  the same order of magnitude.
\par

\section{Concluding Remarks}
\label{sec4}

 \medskip

In this paper, we have investigated the properties of  the plasmon spectra for a
heterostructure consisting of a pair of identical  semi-infinite  conductors and a 2D
graphene  layer sandwiched between them.  Our formulation is suitable when the separation
between the two semi-infinite bulk conducting  materials is small  compared to the inverse
Fermi wave number in graphene,   and the whole system is symmetric
in the $z-$direction  perpendicular to the 2D layer.

 \medskip
\par

We have obtained the plasmon dispersion relations for this encapsulated graphene system
for  both zero and finite energy bandgap and for various values of the chemical potential.
In each case, we clearly obtain three hybridized plasmon  modes, one of which (the acoustic branch)
starts from the origin and is attributed primarily to the graphene layer and the other
two modes originating at $\omega=\omega_p$ are considered as optical  plasmons. This situation
is novel and has not been encountered previously in a system involving a 2D layer   with a single
semi-infinite conductor. Each of the branches exhibits  a specific behavior depending on
the chemical potential and consists of various undamped parts which are determined by the
energy bandgap and its ratio to the doping parameter, as discussed above.

\medskip
\par

The low-frequency branch  has a linear dispersion at long wavelengths and becomes damped
by the intra-band and inter-band particle-hole modes as  wave vector is increased. There
 are two bulk  plasmon branches which are depolarization  shifted by the Coulomb interaction.
 In this regard,  the results demonstrate that there is almost no dependence on the distance
$a$ between the two substrates as long as the condition $a \ll k_F^{-1}$ is satisfied.
Moreover, there is no evidence of
critical   damping of the acoustic plasmon branch in the long wavelength limit
  ($q_\parallel \rightarrow 0$), as  was found  in the previous   comparative study of a
  monolayer of graphene interacting with a single  conducting substrate  \cite{GG}.  Furthermore,
  some crucial properties of the plasmons in free-standing graphene with an energy bandgap,
  such as extension of the undamped branch and its separation into the two parts for intermediate
  energy gap  $\Delta \backsimeq 0.22 \mu$, are also present in our results.

\medskip
\par
 In summary, we have developed  a new analytical model  and obtained  a complete set of numerical results for
a two-dimensional layer (graphene) surrounded by two identical thick conducting substrates. While our
previous work \cite{GG} confirmed and offered an adequate theoretical explanation for  recent experimental
findings \cite{Pol1,Pol2,Pol3,Pol4}, this paper is expected to predict the correct plasmon behavior
of the totally realistic and novel situation of encapsulated graphene, which is now being very
actively studied experimentally \cite{Gong,Kamat}. Hybridized plasmon modes of a graphene-based nanoscale
system are at the focus  of significant interest in the  current fields of technology and applications.
The authors of \cite{encaps5} report evidence of the formation of electron-hole puddles
for encapsulated graphene by hexagonal BN. This is an interesting effect which was not taken into account
in our model calculations. These localization effects will be investigated in future work by including a
concentration of impurities and defects in interaction with the 2D layer and two substrates.
The role of impurities is expected to be modest at low concentration. An increase in the  concentration
of the impurities  will lead to a decrease in the electron correlations. The Green's function and
consequently will, in this case, be  expressible by vertex corrections by the usual rules
of field theory when impurities are present.

%------------------------------


\begin{references}


\bibitem{encaps1}  Principi A., Carrega M., Lundeberg M. B., Woessner A.,
Koppens F. H. L., Vignale G., and Polini M.,``Plasmon losses due to electron-phonon scattering: The case of graphene encapsulated in hexagonal boron nitride,"
Phys. Rev. B {\bf 90}, 165408 (2014).


\bibitem{encaps2}  Kharche N. and Nayak S.K.,``Quasiparticle Band Gap Engineering of Graphene and Graphone on Hexagonal Boron Nitride Substrate," Nano Lett.   {\bf  11}, 5274 (2011).


\bibitem{encaps3}  Ristein J., Mammadov S., and Seyller T,``Origin of Doping in Quasi-Free-Standing Graphene on Silicon Carbide," \prl {\bf 108}, 246104 (2012).


\bibitem{encaps4}  Ryzhii V.  and Satou A.,``Plasma waves in two-dimensional electron-hole system in gated graphene heterostructures,"  Jour. Appl. Phys. {\bf 101}, 024509 (2007).

\bibitem{encaps5}  Woessner A.,	Lundeberg M. B.,	Gao Y.,	Principi A.,	González P.A., Carrega M.,	Watanabe K., Taniguchi T.,	Vignale G.,	Polini M., Hone J.,	Hillenbrand R., and Koppens F.H.L.,``Highly confined low-loss plasmons in graphene–boron nitride heterostructures,"  Nature Materials {\bf 14}, 421  (2015).


\bibitem{encaps6} Mayorov A. S., Gorbachev R. V., Morozov S.V., Britnell L., Jalil R., Ponomarenko L.A.,Blake B., Novoselov K.S., Watanabe K., Taniguchi T., and Geim A.K.,``Micrometer-Scale Ballistic Transport in Encapsulated Graphene at Room Temperature,"
    Nano Lett. {\bf 11} (6),   2396 (2011).

\bibitem{encaps7} Guimarães M.H.D., Zomer P.J., Ingla-Aynés J., Brant J.C., Tombros N., and Wees
B.J.V.,``Controlling Spin Relaxation in Hexagonal BN-Encapsulated Graphene with a Transverse Electric Field," Phys. Rev. Lett. { \bf 113} 086602 (2014).


\bibitem{encaps8}    Yang S.,  Feng X.,  Ivanovici S.,and Müllen K., ``Fabrication of Graphene-Encapsulated Oxide Nanoparticles: Towards High-Performance Anode Materials for Lithium Storage,"
  Angewandte Chemie.  {\bf 49},    8408    (2010).


 \bibitem{encaps9}  Britnell L., Gorbachev R.V., Jalil R., Belle B.D.,
  Schedin F., Mishchenko A., Georgiou T., Katsnelson M.I., Eaves L.,  Morozov5 S.V.,
 Peres N.M.R., Leist J., Geim A.K,  Novoselov K.S., Ponomarenko L.A.,``Field-Effect Tunneling Transistor Based on Vertical Graphene Heterostructures,"
 Science {\bf  335}, 6071  (2012).

   \bibitem{encaps10}   Kretinin A.V.,  Cao Y., Tu J.S., Yu G.L., Jalil R. , Novoselov K.S.,  Haigh S.J.,  Gholinia A.,  Mishchenko A., Lozada M.,  Georgiou T.,  Woods C.R.,  Withers F., Blake P., Eda g.,  Wirsig A.,  Hucho C.,  Watanabe K., Taniguchi T., Geim A.K., and Gorbachev R.V.,``Electronic Properties of Graphene Encapsulated with Different Two-Dimensional Atomic Crystals,"         Nano Lett. {\bf 14 },   3270 (2014).



\bibitem{GG}  Gumbs G,  Iurov A., and Horing N.J.M.,``Nonlocal plasma spectrum of graphene interacting with a thick conductor," \prb {\bf 91}, 235416 (2015).

\bibitem{ONB} Horing N.J.M., Iurov A.,  Gumbs G., Politano A., $\&$ Chiarello G.,``Recent Progress on Nonlocal Graphene/Surface Plasmons," (pp. 205-237), Springer International Publishing (2016).

\bibitem{NJMH}  Horing N. J. M.,``Coupling of graphene and surface plasmons," \prb {\bf 80}, 193401 (2009).

\bibitem{Pol1}   Politano A.,  Marino A.R.,  Formoso V., Farías D.,  Miranda R., and  Chiarello G.,``Evidence for acoustic-like plasmons on epitaxial graphene on Pt(111),"
Phys. Rev. B  {\bf 84}, 033401 (2011).


\bibitem{Pol2}   Politano  A., Marino  A. R.,  and    Chiarello G.,``Effects of a humid environment on the sheet plasmon resonance in epitaxial graphene," Phys. Rev. B {\bf  86},
 085420 (2012).


\bibitem{Pol3}     Politano A.,   Formoso V., and    Chiarello G.,``Evidence of composite
plasmon–phonon modes in the electronic response of epitaxial graphene,", Journal of
Physics: Condensed Matter, {\bf 25} (34), 345303 (2013).


\bibitem{Pol4}   Politano A. and  Chiarello G.,``Quenching of plasmons modes in
air-exposed graphene-Ru contacts for plasmonic devices ,", Applied
Physics Letters, {\bf  102}  , 201608. (2013).


\bibitem{Pol5}    Politano A.   and  Chiarello G.,  ``Unravelling suitable graphene–metal
contacts for graphene-based plasmonic devices,"   Nanoscale,
{\bf 5}, 8215(2013).





\bibitem{DD1} Chuang H. J., Tan  X., Ghimire  N. J.,  Perera M. M., Chamlagain B.,
Cheng  M. M. C.,  Yan J., Mandrus  D.,  Tomanek D., and Zhou Z., ``High Mobility
WSe$_2$  p- and n-Type Field-Effect Transistors Contacted by Highly Doped Graphene
for Low-Resistance Contacts," Nano Lett. {\bf 14}  3594–3601 (2014).


\bibitem{DD2} Shih C.J., Wang Q. H. ,  Son Y.,  Jin Z.,  Blankschtein D., and
Strano M. S., ``Tuning On-Off Current Ration and Field-Effect Mobility in a $MoS_2$
Graphene Heterostructure via Schottky Barrier Modulation," ACS Nano, {\bf 8},
 5790  (2014).

\bibitem{DD3} Low C. G.,  Zhang Q.,   Hao Y., and  Ruoff R. S.,``Graphene Field
Effect Transistors with Mica as Gate Dielectric Layers," Nano Small Micro,
{\bf 10},  4213 ,(2014).

\bibitem{DD4} Kretinin A.  V.,   Cao Y.,  Tu J.  S.,   Yu G.  L.,  Jalil  R.,
Novoselov  K.  S. ,   Haigh S.  J.,   Gholinia A.,   Mishchenko A.,   Lozada M.,
 Georgiou  T.,   Woods C.  R.,  Withers F. ,   Blake P.,   Eda G., Wirsig  A.,
Hucho  C.,   Watanabe K.,  Taniguchi  T., Geim  A.  K. , and    Gorbachev R.  V.,
``Electronic Properties of Graphene Encapsulated with Different Two-Dimensional
Atomic Crystals," Nano Lett.  {\bf  14}, 3270 (2014).

 \bibitem{DD5} Sun T.,  Wang Z. L., Shi  Z. J., Ran  G. Z. ,  Xu W. J., Wang Z. Y. ,
Li Y. Z.,  Dai L., and  Qin G. G.,``Multilayered graphene used as anode of organic
light emitting devices," Applied  Phys. Lett. {\bf 96}, 133301 (2010).

 \bibitem{DD6}  Dean C.R.,  Young A.F., Meric  I.,  Lee C., Wang  L., Sorgenfrei S. ,
Watanabe K.,  Taniguchi T.,  Kim P.,  Shepard K. L., and  Hone J. ``Boron nitride
substrates for high quality graphene electron
ics,"	    Nature Nanotechnology {\bf 5}, 722 (2010).


\bibitem{EGUI}   Eguiluz, A.  Ying S. C., and Quinn J. J.,
``Influence of the electron density profile on surface plasmons in a hydrodynamic model",
Phys. Rev. B {\bf 11}, 2118 (1975).

\bibitem{Kamen}  Horing N. J. M., Kamen  E. and, Gumbs  G.,
``Surface correlation energy and the hydrodynamic model
of dynamic, nonlocal bounded plasma response," Phys. Rev. B
{\bf 31}, 8269 (1985).



\bibitem{Wunsch}    Wunsch B.,  Stauber T., Sols  F., and Guinea  F.,``Dynamical
 polarization of graphene at finite doping," New Jour. of Phys.   {\bf 8}, 318 (2006).

\bibitem{pavlo}  Pyatkovskiy  P.  K.,``Dynamical polarization, screening, and plasmons in gapped graphene,"  J. Phys.: Condens. Matter {\bf 21}, 025506 (2009).

\bibitem{NPo1}   Politano A.,  Silkin V.M.,  Nechaev I.A.,  Vitiello M.S.,  Viti L., Aliev  Z.S.,  Babanly M.B., Chiarello G. ,  Echenique P.M., and  Chulkov E.V.,``Interplay of Surface and Dirac Plasmons in Topological Insulators: The Case of $Bi_2Se_3$,"   Phys. Rev. Lett.{ \bf 115} 216802 (2015) .

\bibitem{NPo2}  Kepaptsoglou D. M., Gilks  D.,  Lari L., Ramasse Q. M.,  Galindo P., Weinert M.,  Li L.,  Nicotra G. and Lazarov  V. K.,``STEM and EELS study of the
Graphene $Bi_2Se_3$ Interface,"
Microsc. Microanal. {\bf 21} 1151 (2015).

\bibitem{Gong}  Gong C.,  Hinojos D., Wang  W.,  Nijem N.,  Shan B.,  Wallace R. M., Cho K., and  Chabal Y. J.,``Metal–Graphene–Metal Sandwich Contacts for Enhanced Interface Bonding and Work Function Control"
ACS Nano {\bf 6}, 5381–5387  (2012).

\bibitem{Kamat}  Kamat P. V.,``Graphene-Based Nanoarchitectures. Anchoring Semiconductor and Metal Nanoparticles on a Two-Dimensional Carbon Support," J. Phys. Chem. Lett., {\bf  1} (2), 520–527 (2010).




\end{references}
\end{document}